\documentclass[reprint,amsmath,amssymb,amsfonts,aps,prb,floatfix,]{revtex4-2}
\usepackage{times}
\usepackage{graphicx}
\usepackage{dcolumn}
\usepackage{bm}
\usepackage{ulem}
\usepackage[colorlinks=true, citecolor=blue, linkcolor=blue, urlcolor=blue]{hyperref}
\usepackage{bbold}
\usepackage{wasysym}
\usepackage[usenames,dvipsnames]{xcolor}

\begin{document}

\title{
Domain formation driven by the entropy of topological edge-modes
}
\author{Gal Shavit}
\author{Yuval Oreg}
\affiliation{
 Department of Condensed Matter Physics, Weizmann Institute of Science, Rehovot, Israel 76100
}

\date{\today}

\begin{abstract}
In this manuscript we study interacting systems with spontaneous discrete symmetry breaking, where the degenerate symmetry-broken states are topologically distinct gapped phases. Edge modes appear at domain walls between the two topological phases. 
In the presence of a weak  disorder field conjugate to the order parameter, we find that the entropy of the edge-modes drives a thermal transition between a gapped uniform phase and a phase with a disorder-induced domain structure.
We characterize this transition using a phenomenological Landau functional, and corroborate our conclusions with a concrete microscopic model.
Finally, we discuss the possibilities of experimental signatures of this phase transition, and propose graphene-based moir\'e heterostructures as candidate materials in which such a phase transition can be detected.
\end{abstract}

\maketitle
\textit{Introduction.---}
Topological phases of matter ~\cite{HasanKaneRMPtopology,QiZhangRMPtopology,bernevigbook} have been a subject of great interest in condensed matter physics ever since the discovery of the integer and fractional quantum Hall effects~\cite{KlitzingIQH,TsuiFQH}.
In such systems, there exists a bulk invariant, e.g., the Chern number, distinguishing different phases of matter with the same symmetry, in contrast to Landau's approach to phase transitions, predicated on symmetry-breaking~\cite{Landau:480041}.
In the Chern insulator~\cite{HaldaneModel1988} and topological insulator phases~\cite{BHZ_QSH}, one finds a remarkable feature:  protected metallic edge-modes emerge at the boundary between topologically-distinct phases.

In materials where the electron-electron interactions are non-negligible, the phase diagram often features correlation-driven spontaneously-symmetry-broken phases.
Prominent examples include quantum Hall ferromagnetism~\cite{bilayerQHFM,GrapheneQHFM}, Wigner crystals~\cite{WignerCrystal}, charge and spin density waves~\cite{CDWevidence,SDWstudy}, and metal-insulator transitions~\cite{RMPmetalinsulatortransitions}.

\begin{figure}
\begin{centering}
\includegraphics[scale=0.33]{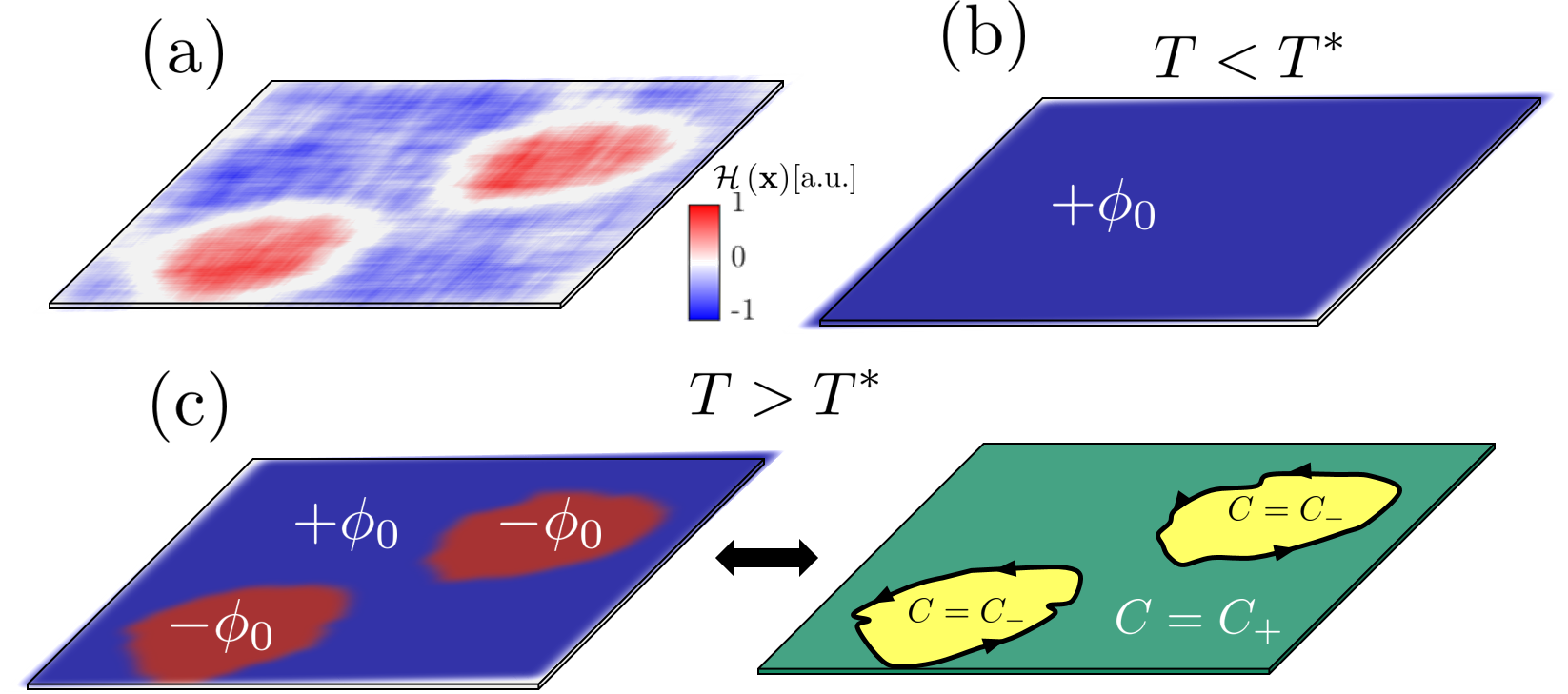}
\par\end{centering}
\caption{ \label{fig:schematics} 
Schematic description of our main results.
In a strongly-correlated two-dimensional system with discrete $\mathbb{Z}_2$ symmetry, electrons may spontaneously break the symmetry and condense into one of two phases characterized  an order parameter $\phi=\pm \phi_0$. Here we study the case where these phases are \textit{topologically distinct}, so that their Chern numbers are different.
We consider  a system subjected to a weak disorder field~${\cal H}\left(\mathbf{x}\right)$, which linearly couples to the order parameter [Eq.~\eqref{eq:disorderF}].
(a) A particular disorder field realization.
(b)
At low enough temperatures the energetically-favored state is uniform, as the system spontaneously chooses $+\phi_0$ (blue).
(c)
At temperatures above $T^*$, determined by both the disorder and the topology [Eq.~\eqref{eq:Tstar}], the system forms domains of the degenerate states $\pm\phi_0$ (blue, red). 
The transition occurs due to the entropy of DW edge-modes, which occur due the different Chern numbers of the gapped $\pm\phi_0$ phases (right panel).
The domain structure is determined by the disorder realization.
}
\end{figure}

In this manuscript, we study how the underlying topological nature of the strongly-interacting electrons may affect their phase diagram.
We focus our interest in systems where the broken symmetry is discrete, and the degenerate symmetry-broken states are topologically inequivalent.
In the presence of disorder which couples to the Ising-like order parameter, the entropy of the metallic edge-modes at domain walls (DWs) drives a thermal phase transition from a uniform phase to a ``domain phase'' whose spatial shape is determined by the disorder landscape (see Fig.~\ref{fig:schematics}).
Our theory may be relevant to quantum Hall plateau transitions~\cite{ChalkerIQH}, class-D superconductors~\cite{ClassDsuperconductors},
the fractional quantum Hall state at filling factor $\nu=5/2$~\cite{Pfaffiandisorder,AshvinHalperin5/2,thermalHallQuantization},
quantum Hall ferromagnetism in graphene~\cite{KharitonovGraphene},
and Chern insulators in graphene moir\'e heterostructures~\cite{AHEprlBultnickZalatel,BilayerMonolayerChern2,ThomsonAlicea}.

Employing field theoretical considerations, we characterize the dependence of this transition on temperature, disorder, and the topological properties of the symmetry-broken states.
We demonstrate our findings using a concrete microscopic model of a two-flavor Chern insulator with strong on-site repulsion.
The dependence of the phase transition on temperature and the strength of the disorder field in the microscopic model is found to be in remarkable agreement with the field-theoretical considerations, verifying the role of the edge-modes entropy at the DWs.

We propose that graphene moir\'e heterostructures~\cite{CaoCorrelatedInsulator,CaoUnconventionalSC,hBNgoldhaberGordon,hBNyoung,TwistedDoubleBilayerYankowitz,TwistedDoubleBilayerShen2020} may be a promising platform to observe and study this effect.
These materials combine the unique topological properties of graphene multi-layers, and effectively large Coulomb repulsion in the flat bands, making them ideal candidates for the physics we discuss.
Finally, we explain how the phase transition of interest can be experimentally measured with currently accessible high-resolution probes.

\textit{Theory.---}
We consider a spontaneous-symmetry-breaking phase transition described by the Ising-order Landau free-energy functional
\begin{equation}
    F_{\rm L}=\int d^{2}\mathbf{x}\left[-\frac{r}{2}\phi^{2}+\frac{c}{2}\left(\nabla\phi\right)^{2}+u\phi^{4}\right], \label{eq:GeneralF}
\end{equation}
where $\phi$ is a real scalar field, and $r$, $c$, and $u$ are positive. We denote $f_0=\frac{r^2}{16u}$ as the condensation energy density for the uniform solution, $\phi\left({\mathbf{x}}\right)=\pm\phi_0=\pm\sqrt{\frac{r}{4u}}$. Notice $F_{\rm L}$ possesses $\mathbb{Z}_2$  symmetry, $\phi\to-\phi$.

Let us examine a DW configuration along the $x$ direction with the ansatz $\phi\left({\mathbf{x}}\right)=\phi_0\tanh\frac{x}{\xi}$, where $\xi$ is a correlation length determined by minimization the DW free-energy. We may then find the DW energy per unit length $L$, or surface tension, $J_{\rm DW} = \frac{8}{3}f_0\xi$,
and $\xi=\sqrt{2c/r}$. 

We introduce disorder which couples linearly to $\phi$ in the spirit of Ref.~\cite{ImryMa}, $F_{\rm L}\to F_{\rm L}+F_{\rm dis}$, with
\begin{equation}
    F_{\rm dis}=\int d^{2}\mathbf{x}{\cal H}\left(\mathbf{x}\right)\phi\left(\mathbf{x}\right),\label{eq:disorderF}
\end{equation}
and for concreteness, we consider Gaussian-correlated disorder, i.e., ${\cal H}$ has zero mean and its variance is
\begin{equation}
    \overline{{\cal H}\left(\mathbf{x}\right){\cal H}\left(\mathbf{x'}\right)}=h^2\exp\left[-\frac{\left(\mathbf{x}-\mathbf{x'}\right)^2}{2\lambda^2}\right].\label{eq:gaussianvariance}
\end{equation}
As a consequence, a judiciously chosen domain of linear size $L$ will benefit from an energy reduction of $ F_{\rm dis}\approx - h \lambda L\sqrt{1-e^{-L^2/2\lambda^2}}$,
which scales as $\sim-h\lambda L$ when the domain size is $\apprge \lambda$.
Considering short-range disorder instead, our conclusions remain unchanged~\cite{SM}.

Let us define the ratio between the energy gain of a domain due to the disorder and the energy cost of the DW, as the \textit{dimensionless disorder strength},
\begin{equation}
    \alpha\equiv \frac{F_{\rm dis}}{J_{\rm DW}L}=\frac{3}{8}\frac{h\lambda}{f_0 \xi}.\label{eq:dimensionlessdisorder}
\end{equation}
Notice the expression for $\alpha$ contains both the disorder magnitude $h$ and the disorder correlation length $\lambda$.

One would normally expect that the condition for favoring the domains configuration is $\alpha>1$
~\footnote{When domain-roughening effects are considered~\cite{Binder1983}, The critical value of $\alpha$ in a finite system is smaller than 1, see SM~\cite{SM}}. 
In this work we demonstrate how the topological properties of the system may alter this condition. Specifically, we show that at finite temperatures $\alpha<1$ may be sufficient for the system to transition into a domain-patterned phase, due to the edge-modes entropy.

Consider now the case where the two degenerate solutions with order parameter $\pm \phi_0$ are topological insulators with different Chern numbers, $C_\pm$ respectively. One expects from the bulk-edge correspondence that a DW between these phases hosts $N_{\rm ch}$ topologically-protected chiral edge-modes, with $N_{\rm ch}=\left|C_+-C_-\right|$. 
These modes will have an entropic contribution to the free-energy of the domains configuration at finite temperature, $F_{\rm ent}=-TsL$, where the entropy per unit length of each edge-mode with velocity $v$ is~\cite{SM}
\begin{equation}
    s=\frac{\pi^2}{3}\frac{k_B^2 T}{\hbar v}.\label{eq:entropyofT}
\end{equation}

We now find the transition point from a uniform $\left|\phi\right|=\phi_0$ solution, to a configuration with domains following the topography of the disorder, with domain density $\propto \lambda^{-2}$. The transition is found by equating the total free-energy of both configurations, i.e., solving
\begin{equation}
    J_{\rm DW}\left(1-\alpha\right)-N_{\rm ch}Ts=0,\label{eq:transitionequation}
\end{equation}
from which we may recover the transition temperature $T^*$,
\begin{equation}
    T^*=T_0\sqrt{\left(1-\alpha\right)},\,\,\,\,\,k_B T_0=N_{\rm ch}^{-1/2}\sqrt{\frac{8}{\pi^2} f_0 \xi^2 \frac{\hbar  v}{\xi}}\label{eq:Tstar}.
\end{equation}
The temperature $T_0$ is an emergent energy scale, above which \textit{spontaneous} DW formation is favored in a disorder-free system ($\alpha=0$).
In order for such a transition to be observable in the absence of disorder, both energy scales $f_0 \xi^2$ and $ \hbar v / \xi$ must be sufficiently small as compared to the condensation temperature of $\phi$.
However, in the presence of disorder the transition to the non-uniform case may occur at temperature much smaller than $T_0$ or the condensation temperature.

Interestingly, the transition temperature is driven \textit{down} by a larger topological disparity between the phases, as it leads to more protected edge-modes which lower the free-energy of the domain phase via their entropy. 
Notice the separability of $T^*$, where one part of it is determined by the disorder strength $\alpha$ and independent of topology, whereas the other part $T_0$ is unaffected by disorder and is determined by the energetics and topology of the condensed phase.

The effect of an additional competing phase, which does not couple to the disorder field, and its possible relevance to magic-angle twisted bilayer graphene (MATBG), is explored in the supplementary materials (SM)~\cite{SM}.

Naturally, in a given system and disorder realization, the strength of the disorder field and size of the characteristic domains may vary around their nominal values of $h$ and $\lambda$, respectively.
In that case, one expects the phase transition at $T^*$ to be broadened: beginning in the uniform $\pm\phi_0$ phase, increasing the temperature will not necessarily ``flip'' all the domains in the system at once.
Examining Eq.~\eqref{eq:dimensionlessdisorder}, implies that larger stronger domains will flip first, followed by the weaker smaller ones.
The width of the system-wide transition as a function of temperature will thus be determined by the variance of $h\lambda$ throughout the sample.

In the next section, we will focus on the transition of an individual domain in a microscopic model, and show how it relates to the results obtained above.

\begin{figure}
\begin{centering}
\includegraphics[scale=0.42]{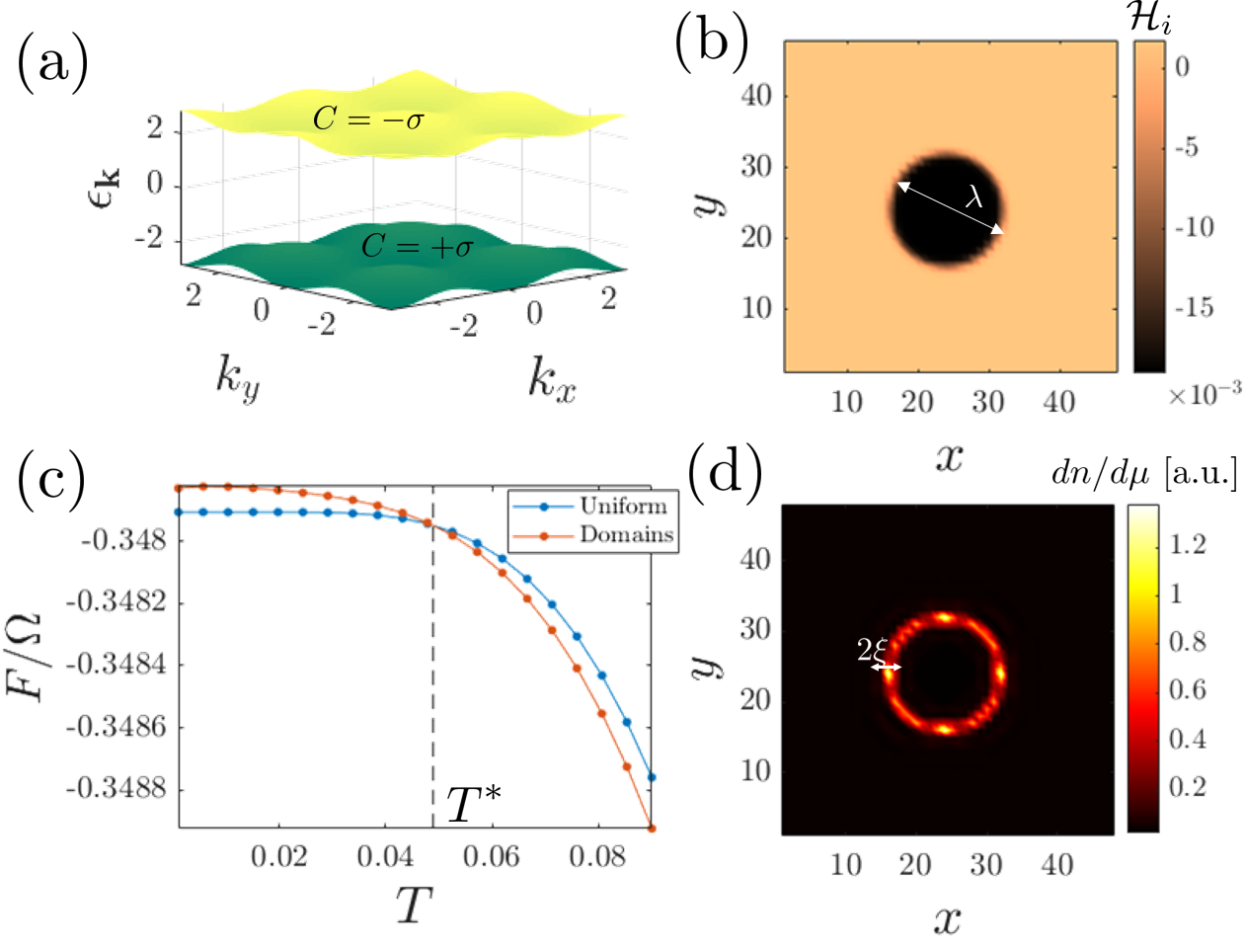}
\par\end{centering}
\caption{\label{fig:modelresults} 
(a)
Non-interacting spectrum of the microscopic lattice model.
For each species $\sigma$, the conduction and valence bands have opposite Chern numbers.
The spectrum shown is doubly degenerate, where the same bands in different species have opposite Chern numbers.
(b)
An example of a specific realization of the field ${\cal H}_i$ which couples to the species polarization [Eq.~\eqref{eq:Hdisorder}].
The characteristic domain correlation length $\lambda\sim 16$ sites is indicated. 
Notice $\sum_i{\cal H}_i=0$.
(c)
Free-energy per site of the self-consistent mean-field solution, where the initial conditions are either uniform or with a domain structure. A dashed vertical line  marks the transition temperature $T^*$.
(d)
Local compressibility at $T=0.062>T^*$. We find compressible regions at the domain boundaries between two regions with differing Chern numbers.
The width of the DW, which is roughly the correlation length $\xi\sim 1$ site is indicated.
In (c)--(d) we use $U=2.5$ and the particular realization shown in (b). Calculations were done on a $48\times48$ lattice with periodic boundary conditions. Energy units are set by the nearest-neighbor-hopping $t_1=1$, see Eq.~(\ref{eq:Hamiltonian}).
}
\end{figure}

\textit{Microscopic model.---}
We consider two species of electrons on a square lattice, described by annihilation operators at site $i$, $c_{\sigma,i}$, with $\sigma=\pm$ for different species, with on-site interaction $U$,
\begin{equation}
\label{eq:Hamiltonian}
    H = \sum_{\sigma,i,j}c_{\sigma,i}^\dagger t_{ij}^\sigma c_{\sigma,j} + U\sum_i n_{+,i}n_{-,i},   
\end{equation}
and $n_{\sigma,i}\equiv c_{\sigma,i}^\dagger c_{\sigma,i}$.
We use a model introduced in Ref.~\cite{titusFQHmodel} for $t_{ij}^\sigma$, with nearest-neighbor-hopping strength set to $t_1=1$ (henceforth setting the energy units) and next-nearest-neighbor hopping $t_2=1/\sqrt{2}$.
The non-interacting part features two energy bands per species, separated by a large gap, see Fig.~\ref{fig:modelresults}a. The upper and lower bands have opposite Chern numbers $C=\pm 1$, and we employ $t_{ij}^+=\left(t_{ij}^-\right)^*$, which leads to opposite Chern numbers for the same band in different species. The Hamiltonian thus has a $\mathbb{Z}_2$ symmetry, corresponding to swapping of the species and complex conjugation (exchanging the Chern numbers).

We consider this system at quarter-filling, where we find that the on-site repulsion may lead to spontaneous breaking of the $\mathbb{Z}_2$ symmetry due to a mean-field Stoner-like instability, and the development of species polarization.
If $U$ is large enough, a species-polarized Chern insulator forms, where the lower-energy band of one species is fully occupied, and all the other bands are completely empty~\cite{SM}.

We now introduce a Zeeman-like term which couples to the $\mathbb{Z}_2$ order parameter, $H\to H+H_{\rm dis}$, with
\begin{equation}
    H_{\rm dis}=\sum_i {\cal{H}}_i\left(n_{+,i}-n_{-,i}\right).\label{eq:Hdisorder}
\end{equation}
For simplicity we consider the case where ${\cal H}_i$ is negative in a single large domain and  positive elsewhere with a zero mean $\sum_i{\cal H}_i=0$, see Fig.~\ref{fig:modelresults}b. 
This may be thought of as a ``zoom-in'' on a particular domain of some specific disorder realization in a much larger system.
The parameter $\lambda$ used above to characterize the disorder correlation length is approximately the diameter of the domain.

We now employ a self-consistent spatial-dependent mean-field approach, approximating
$U n_{+,i}n_{-,i} \approx U \left\langle n_{+,i}\right\rangle n_{-,i}+U n_{+,i}\left\langle n_{-,i}\right\rangle-U \left\langle n_{+,i}\right\rangle \left\langle n_{-,i}\right\rangle$.
We calculate $\left<n_{\sigma,i}\right>$ at different temperatures and magnitudes of $H_{\rm dis}$.
We note that throughout this work our real-space calculations are done with periodic boundary conditions, in order to exclude the entropic effects of edge-modes on the system boundaries.
For more details on the microscopic model and mean-field calculations, see SM~\cite{SM}.

In Fig.~\ref{fig:modelresults}c we compare the free-energy $F$ for a self-consistent solution, where the initial~$n_{\sigma,i}$ we plug into the iterative process are either (i) roughly uniform and the system globally has the same species polarization, or (ii) the polarization within the domain is opposite to that outside of it~\cite{SM}. Both choices yield a stable self-consistent solution, reflecting the existence of local minima of the free-energy.
In the non-uniform scenario, the Chern number inside the domain is opposite to the one outside, and one expects to find chiral edge-modes on the domain boundaries. 
The transition between the uniform and the non-uniform phases is the focus of our work.

As we expected from our theoretical considerations, at low temperatures the system prefers to avoid the exchange energy costs of having a DW, whereas at high enough temperatures the domain structure is preferred, as one gains entropic free-energy from the topological edge-modes.
To show the existence of these metallic DWs, we compute the local compressibility  for a temperature higher than the transition temperature $T^*$, cf. Fig.~\ref{fig:modelresults}d. It features a clear compressible ring encircling the negative ${\cal H}_i$ region, while everywhere else the system is in an incompressible Chern insulator state.

\begin{figure}
\begin{centering}
\includegraphics[scale=0.48]{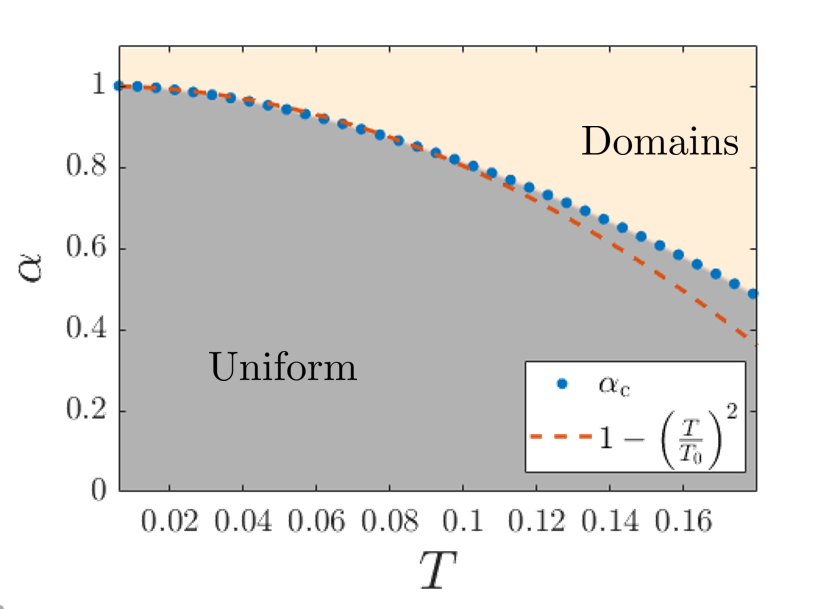}
\par\end{centering}
\caption{\label{fig:phasediagram} 
Phase diagram of the microscopic model at quarter filling.
At each temperature, the critical dimensionless disorder strength $\alpha_c$ is recovered (blue).
As a guide to the eye, we plot $\alpha_c=1-\left(T/T_0\right)^2$ (dashed red line), with $T_0\approx0.22$.
We have defined $\alpha\equiv h/h^0_c$, with $h^0_c$ the critical field RMS at zero temperature.
The $T^2$ dependence is indicative of metallic DW modes lowering the free-energy via their entropy, in agreement with our theoretical prediction [Eq.~\eqref{eq:Tstar}].
In this figure we used the same parameters and the same $H_{\rm dis}$ realization as in Fig.~\ref{fig:modelresults}, yet with scaling of ${\cal H}_i$ to vary its strength.
}
\end{figure}

By varying the strength of $H_{\rm dis}$ via its root-mean-square~(RMS), 
$h=\sqrt{\frac{1}{\Omega}\sum_i {\cal H}_i^2}$, ($\Omega$ is the system volume) and monitoring the evolution of $T^*$, we can compare the microscopic model analysis to our field-theoretical considerations above. 
As shown in Fig.~\ref{fig:phasediagram}, at higher temperatures the critical dimensionless disorder strength is lowered as $\alpha_c\approx 1-\left(T/T_0\right)^2$, in agreement with the functional form predicted by Eq.~\eqref{eq:Tstar}. 
The $T^2$ dependence is evidence of the entropic contribution of the DW modes towards lowering the free-energy, as their entropy $s$ is approximately linear in $T$.
In the absence of this topological effect one would expect the usual $\alpha_c=1$.

We note that at high enough temperatures the critical disorder strength begins to deviate from this theoretical formula.
A possible explanation for that is the saturation of edge-mode entropy~\cite{SM}. 
As the temperature increases and becomes comparable to the mean-field gap, the linear-in-$T$ form of the entropy [Eq.~\eqref{eq:entropyofT}] no longer holds.
This form assumes a constant-in-energy density of states of the edge-modes, an assumption that loses its validity at energies comparable to the gap, where the edge-modes ``merge'' into the bulk dispersion.

\textit{Experimental consequences.---}
The ingredients needed to manifest the sort of phase transition discussed in this work are (i) topologically non-trivial bands, and (ii) an instability towards spontaneous symmetry breaking due to strong interactions.
Natural candidates for systems having both ingredients are the multi-layer graphene moir\'e heterostructures~\cite{CaoCorrelatedInsulator,CaoUnconventionalSC,hBNgoldhaberGordon,hBNyoung,TwistedDoubleBilayerYankowitz,TwistedDoubleBilayerShen2020,PabloTrilayer}.
In such materials, the interplay between the effectively large interaction due to the flat-band structure and the Berry curvature, may indeed result in phases where the ground state is a ``flavor-polarized'' Chern insulator. 

For example, the microscopic model we explored resembles a spin-polarized version of MATBG aligned with hexagonal Boron-Nitride (hBN)~\cite{AHEprlBultnickZalatel}. 
In that context, the $\pm$ species correspond to the two valleys, and the gap in the non-interacting band structure is a result of $C_2$ symmetry breaking by the hBN substrate.
It is strongly believed that the mechanism that leads to a ferromagnetic Chern insulator in hBN-aligned MATBG \cite{hBNgoldhaberGordon,hBNyoung} is similar in nature to the mechanism present in the model we explored near quarter-filling~\cite{AHEprlBultnickZalatel}.

Another, perhaps more promising candidate system is twisted monolayer-bilayer graphene, where Chern insulators have been observed at odd fillings~\cite{BilayerMonolayerChern2}. The relevant band at the appropriate range of perpendicular electric field (which tunes both the topology of the bands and their width) has Chern number $C=\pm2$. This higher Chern number may lead to  $N_{\rm ch}=4$ in domains with opposite valley polarization, and thus an enhancement of the edge-modes entropy.
Ultimately, enhancement of $N_{\rm ch}$ will contribute to expand the range of disorder regimes where thermal switching between a uniform and a domain-patterned phase can take place. 

Yet, the effect we describe is difficult to uncover using global transport measurements.
One possibility is to use spatially-resolved scanning tunneling microscopy to detect the compressible edge-modes~\cite{DrozdovYazdaniSTM1dBismuth}.
Another relevant tool is the superconducting quantum interference device (SQUID) utilized in Refs.~\cite{ZeldovMatbgSquidAviram,AndreaYoungSquidDomains,ChernMosaicZeldov}, which has spatial resolution of the order of $\sim$ 100 nm.
In Ref.~\cite{AndreaYoungSquidDomains} the domain structure of a suitable ferromagnetic Chern insulator was in fact measured, revealing that the order parameter indeed couples to some spatial disorder field pinning the domains. 
We postulate that a similar measurement as a function of temperature, perhaps in several devices with varying disorder profiles and magnitudes, may reveal the topology-driven phase transition we discuss.

An alternative probing method is local compressibility measurements~\cite{AmirYacobyLocalCompressibility,DiracRevivals}. At a particularly high resolution, one could map the compressibility of the system as a function of temperature and obtain an image resembling Fig.~\ref{fig:modelresults}d.
However, even at a realistic resolution where the DW cannot be fully resolved, one may detect the temperature-driven phase transition.
At the transition, a previously insulating area, the DW, becomes compressible. 
Taking into consideration the mesoscopic charging energy of the DW of order $e^2/L$, it becomes an effective quantum dot.
This will lead to discrete charging events~\cite{IlaniYacobyQHlocalizedstates} when, e.g., a gate-voltage is tuned, which disappear at the low-temperature uniform incompressible phase.

Detection of the one-dimensional chiral nature of the $T>T^*$ compressible state may be done by applying a small magnetic field perpendicular to the sample. 
This will shift the charging events in gate-voltage due to the spectral flow of the chiral modes encircling the insulating domain.
The shift will then be periodic in flux penetrating the encompassed domain with the periodicity of the flux quantum, $\Phi_0=2\pi\hbar/e$.

\textit{Conclusions.---}
We have unveiled and studied a phase transition  occurring in strongly-correlated topological materials with spontaneous discrete symmetry breaking.
Due to the topological nature of the symmetry-broken phase, metallic edge-modes reside on DWs between topologically-distinct degenerate ground states.
In the presence of disorder, the entropic contribution of the edge-modes to the thermodynamic free-energy may make the free-energy of the non-uniform state lower.
The domain structure of this non-uniform phase will be determined by the topography of the disorder field.

Employing field theoretical considerations, we demonstrated the competition between the exchange energy cost of DWs on one hand, and the free-energy gain due to edge-modes entropy and coupling to disorder on the other.
This enabled us to map out the phase boundary separating the uniform and non-uniform ordered phases as a function of both temperature and disorder strength.
We have verified our predictions using a microscopic model of a Chern insulator with strong interactions.
Our analysis of the microscopic model was able to reproduce the relevant phase boundary that was obtained from the phenomenological field theoretical treatment, emphasizing the role of the edge-mode entropy.

Finally, we discussed possible materials suitable for the realization of this unusual phase transition, with graphene-based moir\'e heterostructures showing the most promise.
Experimental detection of the thermal uniform-to-non-uniform transition was proposed using currently available measurement schemes, e.g., scanning SQUID~\cite{ZeldovMatbgSquidAviram} and local compressibility probes~\cite{AmirYacobyLocalCompressibility,DiracRevivals}.

\begin{acknowledgments}
We acknowledge enlightening discussions with Matan Bocarsly, Sammer Grover, Shahal Ilani, Asaf Rozen, Moshe Shechter, Eli Zeldov, and Uri Zondiner.
This project was partially supported by grants from the ERC under the European Union’s Horizon 2020 research and innovation programme (grant agreements LEGOTOP No. 788715), the DFG (CRC/Transregio 183, EI 519/7-1), the BSF and NSF (2018643), and the ISF Quantum Science and Technology (2074/19).
\end{acknowledgments}

\bibliography{chern}


\begin{widetext}
\section*{
Supplementary Materials for "Domain formation driven by the entropy of topological edge modes"
}

\setcounter{section}{0} \renewcommand{\thesection}{S.\arabic{section}} \setcounter{figure}{0} \renewcommand{\thefigure}{S\arabic{figure}} \setcounter{equation}{0} \renewcommand{\theequation}{S\arabic{equation}}

\section {Domain surface tension}
For the sake of completeness, we bring here the calculation of $J_{\rm DW}$, the energy per unit length of a domain wall, in terms of the parameters appearing in the Landau functional,
\begin{equation}
    F_{\rm L}=\int d^{2}\mathbf{x}\left[-\frac{r}{2}\phi^{2}+\frac{c}{2}\left(\nabla\phi\right)^{2}+u\phi^{4}+f_{0}\right]. \label{eq:GeneralFinsupp}
\end{equation}
Notice we have added a constant energy term $f_0=\frac{r^2}{16u}$ to the functional, this will prove convenient when calculating the energy cost of the domain wall.
Let us examine a domain wall configuration along the $x$ direction with the ansatz $\phi\left({\mathbf{x}}\right)=\phi_0\tanh\frac{x}{\xi}$, where $\xi$ is a correlation length to be determined by minimization the free energy.
Plugging this ansatz into Eq.~\eqref{eq:GeneralFinsupp}, we find
\begin{align}
    F_{\rm L, DW} &= L_y \times 2 f_0 \int dx\left[\frac{c/r}{\xi^{2}}\frac{1}{\cosh^{4}\frac{x}{\xi}}+\frac{1}{2}\frac{1}{\cosh^{4}\frac{x}{\xi}}\right]\nonumber\\
    &=L_y \times \frac{4}{3}f_0\left(2\frac{c/r}{\xi}+\xi\right),
\end{align}
which is minimized by setting $\xi=\sqrt{2c/r}$. Thus, we obtain
\begin{equation}
    J_{\rm DW}=\frac{F_{\rm L, DW}}{L_y} = \frac{8}{3}f_0 \xi.
\end{equation}
This result makes intuitive sense, as the energy cost of a domain wall is due to flipping an area of size $\sim L_y\times\xi$ from the condensed phase to the normal phase, which has an added energy density cost of $f_0$.

\section {Energy gain due to coupling to disorder}
The coupling of the disorder field ${\cal H}\left(\mathbf{x}\right)$ to the order parameter $\phi$ is described by the term
\begin{equation}
    F_{\rm dis}=\int d^{2}\mathbf{x}{\cal H}\left(\mathbf{x}\right)\phi\left(\mathbf{x}\right).\label{eq:disorderFSupp}
\end{equation}
We assume ${\cal H}$ has zero mean and its variance is generally given by
\begin{equation}
    \overline{{\cal H}\left(\mathbf{x}\right){\cal H}\left(\mathbf{x'}\right)}= \frac{1}{\phi_0^2}h^2 g \left(\mathbf{x}-\mathbf{x'}\right).\label{eq:generalvarianceSupp}
\end{equation}
We employ a standard method to estimate the disorder energy gain~\cite{ThomsonAlicea} by calculating the root-mean-square of the energy of an arbitrarily chosen domain $\cal D$,
\begin{equation}
    F_{\rm rms}\approx h \sqrt{\int_{\cal D} d^2 \mathbf{x} \int_{\cal D} d^2 \mathbf{x}' g \left(\mathbf{x}-\mathbf{x'}\right)},
\end{equation}
and the integral $\int_{\cal D}$ signifies integration over $\mathbf{x},\mathbf{x'},$ which are both in the domain region. 

Let us now examine two kinds of disorder. First, we consider short-range disorder, i.e., $g \left(\mathbf{x}-\mathbf{x'}\right)=\lambda_{\rm short}^2\delta^2\left(\mathbf{x}-\mathbf{x'}\right)$.
Then, a \textit{judiciously chosen domain} of linear extent $L$
will benefit from an energy reduction of
\begin{equation}
    F_{\rm dis}\approx - F_{\rm rms} \approx -h \lambda_{\rm short} L.
\end{equation}
Similarly, for gaussian-correlated disorder, 
$g \left(\mathbf{x}-\mathbf{x'}\right)=\exp\left[-\frac{\left(\mathbf{x}-\mathbf{x'}\right)^2}{2\lambda_{\rm gaussian}^2}\right]$, one finds
\begin{equation}
    F_{\rm dis}\approx - h \lambda_{\rm gaussian} L\sqrt{1-e^{-L^2/2\lambda_{\rm gaussian}^2}},
\end{equation}
which is roughly $\sim-h\lambda_{\rm gaussian} L$ when the domain size is $\apprge \lambda_{\rm gaussian}$, as one indeed expects a judiciously chosen domain to be.

\subsection{The effect of domain roughening}
We consider the distortion of the interface between domains, or "domain roughening" ~\cite{Binder1983,SchechterPC2022}.
Incorporating this effect leads to a modification of the energy cost of a domain wall from $J_{\rm DW}L$, to $J_{\rm DW}L \left[1-\frac{\alpha^2}{b}\log\frac{L}{\lambda}\right]$. 
As a reminder, $\alpha$ was defined in the main text as the dimensionless disorder strength, $\alpha= \frac{h\lambda}{J_{\rm DW}}$. The dimensionless number $b$ is an order-unity proportionality constant which depends on microscopic details.
This implies the existence of an \textit{exponentially large length scale}, 
\begin{equation}
    L^* \approx \lambda \exp\left(\frac{b}{\alpha^2}\right),
\end{equation}
above which domain-wall formation is favored for arbitrarily weak disorder.

However, for a given system size with linear dimension $L_{\rm sys}$ smaller then the exponentially large $L^*$ that is not the case.
The main effect will be to reduce the critical dimensionless disorder from $\alpha_{\rm crit.}=1$, to a lower value which may by roughly estimated by solving
\begin{equation}
    \alpha_{\rm crit.} = 1-
    \frac{\alpha_{\rm crit.}^2}{b}
    \log\frac{L_{\rm sys}}{\lambda}.
\end{equation}

Our discussion of domain formation at finite temperatures and our conclusions regarding the phase diagram remain unchanged, once we replace the critical dimensionless disorder to be the new $\alpha_{\rm crit.}$ (instead of 1).

\section{Entropy of chiral one-dimensional electrons}
Considering electrons in one dimension with the dispersion relation
\begin{equation}
    E_k=\hbar v k,
\end{equation}
we find that the density of state per unit length of a \textit{chiral} mode (i.e., half of that of normal electrons) is
\begin{equation}
    {\cal{N}}\left(E\right)=\frac{1}{\hbar v}\equiv{\cal N}.\label{eq:chiraldos}
\end{equation}
The entropy per unit length is 
\begin{equation}
    s=-k_B{\cal{N}}\int^\Delta_{-\Delta} dE\left(E\right)\left[f\log f+\left(1-f\right)\log\left(1-f\right)\right],\label{eq:integralentropy}
\end{equation}
with $f=\left[1+\exp\left( E/T\right)\right]^{-1}$, and the chiral mode extends in energies from $-\Delta$ to $\Delta$ (and has no density of states outside this window). 
At low enough temperatures $T\ll\Delta$, we may extend the limits of integration to infinity and obtain the entropy density of a chiral electronic mode,
\begin{equation}
    s=\frac{\pi^2}{3}\frac{k_B^2 T}{\hbar v}.\label{eq:supplinearentropy}
\end{equation}
However, at intermediate temperatures which are comparable to the gap $\Delta$, the linear-in-$T$ entropy begins to saturate.
This leads to the effect discussed following Fig.~3 in the main text, where at higher temperatures the domain-patterned phase ``loses ground'' to the uniform phase, as compared to what one might expect from an entropy linear in temperature.
The saturation of the entropy with increased temperature is demonstrated in Fig.~\ref{fig:entropysaturates}.

\begin{figure}
\begin{centering}
\includegraphics[scale=0.55]{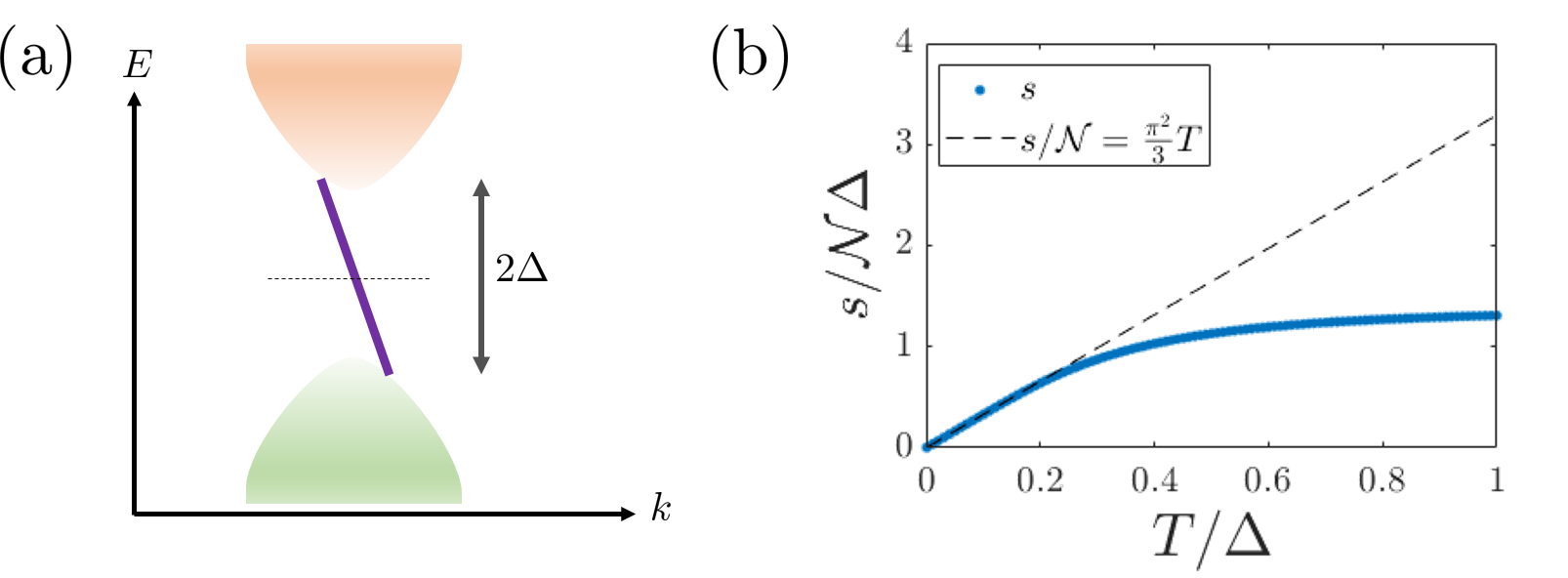}
\par\end{centering}
\caption{\label{fig:entropysaturates} 
(a)
Schematic depiction of a one-dimensional topologically-protected chiral edge mode dispersion. It extends through the bulk gap, in the energy range $E\in\left[-\Delta,\Delta\right]$. 
(b)
Blue dots: entropy per unit length of the chiral edge mode as a function of temperature, extracted from Eq.~\eqref{eq:integralentropy}. Dashed black line: low temperature linear behavior of the entropy [Eq.~\eqref{eq:supplinearentropy}].
As $T$ becomes an appreciable fraction of $\Delta$ the entropy begins to saturate and the deviation from the linear relation grows larger.
}
\end{figure}

\section{Competing phases}
Let us consider a scenario where the  condensed $\phi$-phases are competing in energy with an additional symmetry broken state. We write the free-energy as
\begin{equation}
    F_{\rm comp.}= F_\phi + F_\chi + F_{\phi\chi}+F_{\rm dis},
\end{equation}
with
\begin{equation}
    F_\phi = \int d^{2}\mathbf{x}\left[-\frac{r}{2}\phi^{2}+\frac{c}{2}\left(\nabla\phi\right)^{2}+u\phi^{4}+f_{0}\right],
\end{equation}
\begin{equation}
    F_\chi = \int d^{2}\mathbf{x}\left[-\frac{\tilde{r}}{2}\chi^{2}+\frac{\tilde{c}}{2}\left(\nabla\chi\right)^{2}+\tilde{u}\chi^{4}+f_{\chi}\right],
\end{equation}
\begin{equation}
    F_{\phi\chi} = \int d^{2}\mathbf{x}g\phi^2\chi^2,
\end{equation}
\begin{equation}
    F_{\rm dis}=\int d^{2}\mathbf{x}{\cal H}\left(\mathbf{x}\right)\phi\left(\mathbf{x}\right).
\end{equation}
In the above, $r,\tilde{r},u,\tilde{u}$ are positive, $f_0=\frac{r^2}{16u}$, and $f_\chi=\frac{\tilde{r}^2}{16\tilde{u}}$ are the condensation energies for uniform $\phi$ and $\chi$ phases, respectively.
We assume the energy density $g$ is sufficiently large such that overlapping regions of non-zero $\phi$ and $\chi$ are highly suppressed.
Furthermore, we only consider the $f_\chi>f_\phi$ regime, i.e., a uniform $\chi$ phase is favorable to a uniform $\phi$ phase in the absence of disorder. Notice disorder only couples to one of the sectors, a fact which will play a crucial role.

We want to calculate the transition temperature at which the uniform $\chi$ phase becomes energetically disfavored as compared to the $\phi$-domains phase, which benefits both from the disorder and from the entropy of metallic modes at domain walls. Thus, we need to solve the following equation
\begin{equation}
    -f_\chi\cdot\Omega = -f_0 \cdot\Omega + J_{\rm DW}\cdot L_{\rm tot} - h \lambda \cdot L_{\rm tot} - N_{\rm ch}Ts\cdot L_{\rm tot},\label{eq:competesupp}
\end{equation}
where $\Omega$ is the volume of the system and $L_{\rm tot}$ is the total length of the domain walls in the $\phi$-domain phase. 
We make the reasonable assumption that the volume $\Omega$ is proportional to $\lambda\cdot L_{\rm tot}$, with an order one proportionality constant $z$, which depends on the specific geometry of the domain pattern favored by the disorder realization. Plugging this assumption into Eq.~\eqref{eq:competesupp}, we find the transition temperature
\begin{equation}
    T_\chi^*=T_0\sqrt{\left(1-\alpha+z\frac{\delta}{h}\right)},\label{eq:Tstarchisupp}
\end{equation}
where we have defined the free energy difference $\delta=f_\chi-f_0$. 
At temperatures $T<T_\chi^*$ the system is expected to be in a uniform phase where $\chi$ is condensed and $\phi=0$. At temperatures above the transition, $T>T_\chi^*$, $\chi=0$ throughout the system, and the system will form domains of $\phi=\pm\phi_0$ following the topography of the disorder.

The discussion above is in some sense a generalization of the discussion in Ref.~\cite{ThomsonAlicea}, as we extend it to finite temperatures.
In Ref.~\cite{ThomsonAlicea}, $T_0=0$, and only the disorder parameters and energetics of the competing phases play a role.
The topological nature of (at least) one of the phases was not considered there.

As discussed in Ref.~\cite{ThomsonAlicea}, the competing phases analysis may be relevant to magic angle twisted bilayer graphene (MATBG), where at even integer fillings two competitive phases are likely: (i) a phase with inter-valley coherence (IVC), which is expected to be favored from microscopic considerations \cite{BultnickKhalaf2020,hofmann2021fermionic,ShavitBergSternOreg}, and (ii) a quantum valley Hall state (QVH), with opposite Chern numbers at different valleys.
The QVH state has 2 degenerate phases related by $C_2$ symmetry, which are the analogs of the $\pm \phi_0 $ phases. The disorder considered in Ref.~\cite{ThomsonAlicea} couples directly to the QVH order parameter, yet the IVC phase is decoupled from it (it corresponds to the phase where $\chi$ is condensed).

Our analysis thus reveals the possibility of a scenario for MATBG at even integer fillings, where even in the presence of disorder, at low enough temperature the system is in a uniform IVC phase.
As the temperature increases, switching to the ``domain phase'' of the QVH eventually may then become favorable due to entropy of edge-modes at domain walls.

\section{Details of the mean-field calculations}
In this work we perform calculations using the chiral $\pi$-flux model on a square lattice, see Ref.~\cite{titusFQHmodel}.
The tight-binding Hamiltonian is given in terms of the fermionic annihilation operators $A_{\sigma,i},B_{\sigma,i}$ which annihilate an electron of species $\sigma$ on sublattice $A/B$ of the unit-cell $i$,
\begin{align}
    H_{\pi} &= -t_1\sum_{\sigma,i}
    \left[e^{i\sigma\pi/4}\left( B^\dagger_{\sigma,i-\hat{x}}+B^\dagger_{\sigma,i+\hat{y}} \right)
    +e^{-i\sigma\pi/4}\left( B^\dagger_{\sigma,i}+B^\dagger_{\sigma,i-\hat{x}+\hat{y}} \right)
    \right]A_{\sigma,i} \nonumber\\
    &-t_2\sum_{\sigma,i}\left(A^\dagger_{\sigma,i+\hat{x}}-A^\dagger_{\sigma,i+\hat{y}}\right)A_{\sigma,i}
    +t_2\sum_{\sigma,i}\left(B^\dagger_{\sigma,i+\hat{x}}-B^\dagger_{\sigma,i+\hat{y}}\right)B_{\sigma,i} +{\rm h.c.}.
\end{align}
Throughout this work we use units where $t_1=1$, and we set $t_2=1/\sqrt{2}$.
The model $H_\pi$ has two Chern bands per species separated by a large energy gap (see Fig.~2a in the main text). Opposite species have opposite Chern numbers for a given band, and in each species the conduction and valence bands have opposite Chern numbers $C=\pm 1$ as well.

We introduce on-site interaction energy of the form
\begin{equation}
    H_{\rm int}=U\sum_i\left(
    A^\dagger_{+,i}A_{+,i}A^\dagger_{-,i}A_{-,i}
    +B^\dagger_{+,i}B_{+,i}B^\dagger_{-,i}B_{-,i}
    \right),
\end{equation}
and a possible disorder field coupling to the species polarization,
\begin{equation}
    H_{\rm dis}=\sum_{\sigma,i}2\sigma {\cal H}_i \hat{n}_{\sigma,i},
\end{equation}

where we have defined
$\hat{n}_{\sigma,i}=\frac{1}{2}\left(A^\dagger_{\sigma ,i}A_{\sigma ,i}
    +B^\dagger_{\sigma ,i}B_{\sigma ,i}\right)$.
    
We decompose the interaction Hamiltonian in the density-density channel, and only keep terms up to first order in the deviation from the mean field. The interaction Hamiltonian thus becomes
\begin{equation}
    H_{\rm int,\,MF} \approx 2U\sum_{\sigma,i}\left(n_{\sigma,i}\hat{n}_{\bar{\sigma},i}-\frac{1}{2}n_{\sigma,i}n_{\bar{\sigma},i}\right),
\end{equation}
and $n_{\sigma,i}=\left\langle\hat{n}_{\sigma,i}\right\rangle$ can be found self-consistently from the mean-field Hamiltonian.
This is done by diagonalizing the quadratic mean-field Hamiltonian
\begin{align}
        H_{\rm MF}&=H_\pi+H_{\rm dis}+H_{\rm int,\,MF}-2\mu\sum_{\sigma,i}\hat{n}_{\sigma,i} \nonumber\\
        &=\sum_j\epsilon_j\psi_j^\dagger \psi_j -U\sum_{\sigma,i}n_{\sigma,i}n_{\bar{\sigma},i},
\end{align}
where the last line is written in terms of fermionic operators $\psi_j$ which diagonalize the mean-field Hamiltonian.

Our self-consistent procedure begins with an initial set of occupation numbers, denoted $n_{\sigma,i}^0$. 
At each iteration, we plug $n_{\sigma,i}^j$ into $H_{\rm MF}$, diagonalize it, and calculate $\left\langle\hat{n}_{\sigma,i}\right\rangle$, which constitute the new $n_{\sigma,i}^{j+1}$.
The process is repeated until convergence.

For the sake of completeness, we now present some results for the disorder-free case, $H_{\rm dis}=0$. We use periodic boundary conditions, scan the chemical potential, and self-consistently compute the average occupation number of each species,~ $n_\sigma=\frac{1}{\Omega}\sum_i n_{\sigma,i}$, plotted in Fig.~\ref{fig:uniformfigure}a. 

The species symmetry is visibly broken in large parts of the phase diagram, where one species gets filled, while the other remains inert. Moreover, at $\frac{1}{4}$ and $\frac{3}{4}$ filling, lies a region of incompressibility, which is our regime of interest. For example, in the quarter-filled case, the lower band of $\sigma=+$ is full, while both bands of $\sigma=-$ are empty. The system is thus a Chern insulator with $\left|C\right|=1$.

In the middle of the incompressibility region, we have also calculated the compressibility $dn/d\mu$ as a function of temperature, with $n=\sum_\sigma n_\sigma$.
By fitting the compressibility to be thermally-activated, i.e., $\propto\exp\left(-\Delta/T\right)$, we were able to extract the value of the thermodynamical gap, $\Delta\approx0.21$ (for the interaction strength $U=2.5$).

We comment that the large value of the interaction, which is roughly 3 times that of the bandwidth of the individual Chern bands, is needed in our toy-model in order to induce this gap, and not ``only'' polarization of the species (where $U\sim1$ is sufficient).

The free-energy we used in order to identify the phase transition is calculated from
\begin{align}
    F_{\rm MF} &= -T\sum_j\log\left(1+e^{-\epsilon_j/T}\right) -U\sum_{\sigma,i}n_{\sigma,i}n_{\bar{\sigma},i}\nonumber\\
    &=\sum_j \frac{\epsilon_j-\left|\epsilon_j\right|}{2}-T\sum_j\log\left(1+e^{-\left|\epsilon_j\right|/T}\right)-U\sum_{\sigma,i}n_{\sigma,i}n_{\bar{\sigma},i},
\end{align}
where $n_{\sigma,i}$ were calculated self-consistently. 
We note that the transition to the second line is done to ensure numerical stability of $F_{\rm MF}$ at very low temperatures. At $T=0$ the second term in the second line vanishes, whereas the first term is a simple sum over all negative energies (as appropriate for a ``Fermi sea'').

\begin{figure}
\begin{centering}
\includegraphics[scale=0.6]{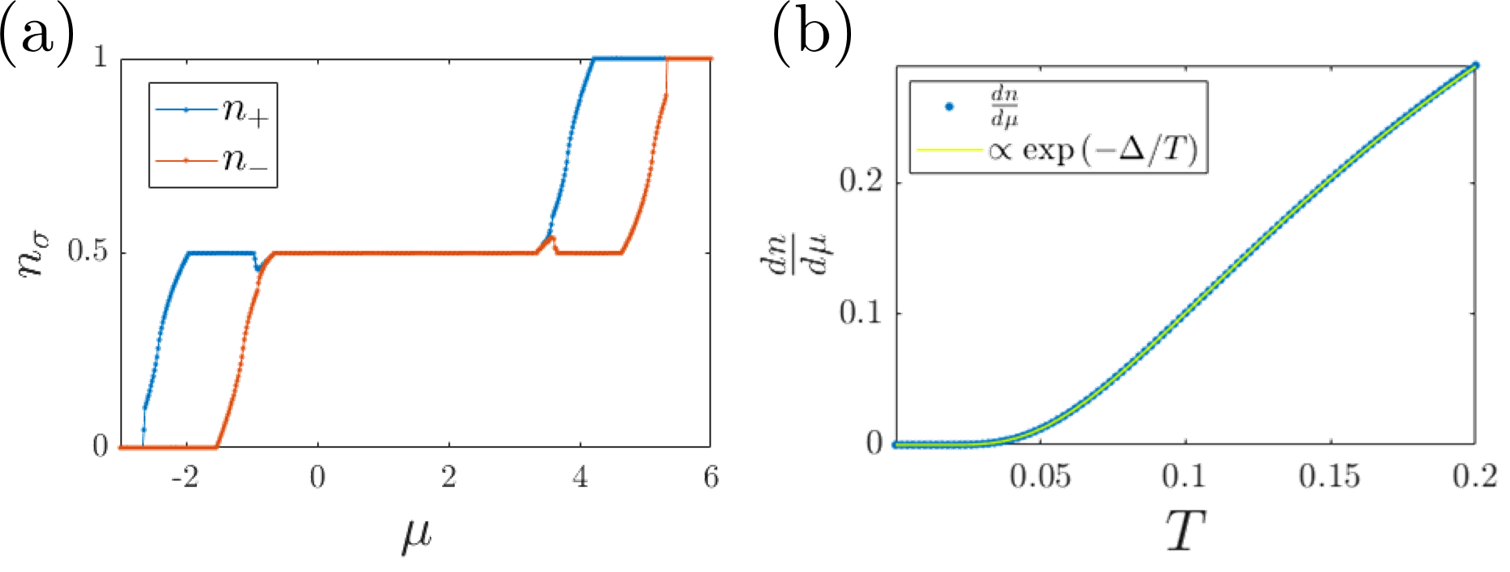}
\par\end{centering}
\caption{\label{fig:uniformfigure} 
Results of the self-consistent mean-field calculations in the clean $H_{\rm dis}=0$ system.
(a)
Average occupation of each species as a function of chemical potential. Due to the large on-site interaction, the ${\cal{Z}}_2$ symmetry is broken at certain fillings, and the system develops spontaneous species polarization. Throughout this work, we work in the quarter-filled regime with $\mu\approx-1.74$, where the system is also gapped and incompressible.
(b)
Temperature dependent compressibility calculated at $\mu=-1.74$. The bright line is an exponential fit to thermally-activated compressibility $dn/d\mu\propto\exp\left(-\Delta/T\right)$, with $\Delta\approx0.21$.
In both panels we used $U=2.5$.
}
\end{figure}

\section{Estimating the correlation length $\xi$}
One of the key parameters in our field-theoretical consideration was the value of the correlation length derived from the Landau free energy functional.
It is interesting to find its value in the microscopic model in the parameter regime we are interested in.
In Fig.~\ref{fig:xi} we plot a cut through the system at a temperature $T>T^*$ of the self consistent mean-field occupation numbers for the two different species, showing the clear opposite-polarization domain formation.
We find that the domain wall is extremely narrow, and by fitting to 
\begin{equation}
    n_{\sigma,x} = \frac{\sigma}{4}
    \left[
    \tanh\frac{x-x_L}{\xi}-\tanh\frac{x-x_R}{\xi}
    \right]
    +\frac{1-\sigma}{4},
\end{equation}
where $x_L$ ($x_R$) is the location of the left (right) domain wall, we find that $\xi\approx 1$ site.
The fact that the correlation length we find is extremely small and close to its minimal possible value is not surprising. This is due to the fact that we are in the strong coupling parameter regime, where the interaction energy scale $U$ is much larger than the bandwidth of the individual bands.

\begin{figure}
\begin{centering}
\includegraphics[scale=0.6]{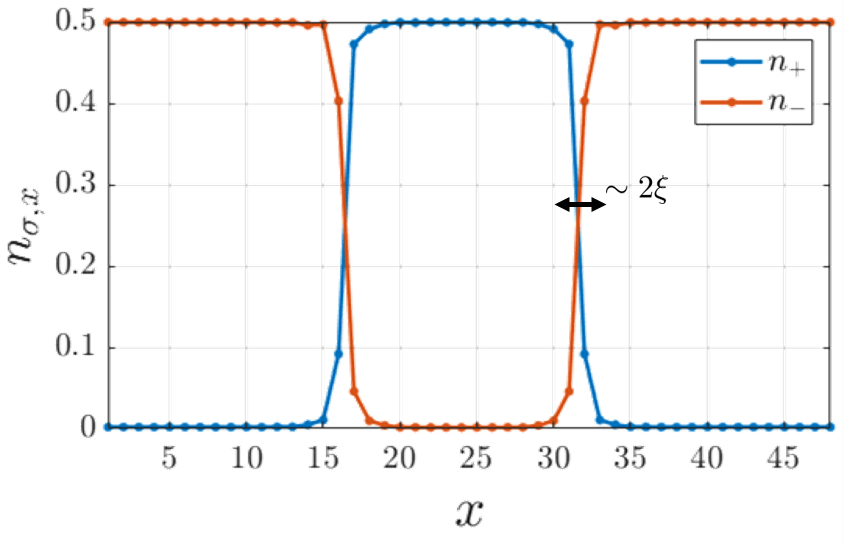}
\par\end{centering}
\caption{\label{fig:xi} 
Estimating $\xi$ in the microscopic model. We plot a cut through the system at $y=24$ of the self-consistent values of the occupation numbers for the two species. The calculation was done on a $48\times48$ lattice with periodic boundary conditions, with the $H_{\rm dis}$ realization appearing in Fig.~2b in the main text, and the parameters $U=2.5$, $T=0.062$.
From this figure, we estimate the size of $\xi$ is roughly 1 site.
}
\end{figure}

\end{widetext}

\end{document}